 \documentstyle{article}
\newfont{\g}{eufm9}

\newcommand{\gtg}{\mbox{\g g}}

\newcommand{\gtb}{\mbox{\g b}}
\newcommand{\hgtb}{\mbox{$\hat{\gtb}$}}
\newcommand{\hgtg}{\mbox{$\hat{\gtg}$}}

\newcommand{\gta}{\mbox{\g a}}
\newcommand{\gtsl}{\mbox{\g sl}}

\newcommand{\nc}{\mbox{${\bf C}$}}

\newcommand{\nz} {\mbox{${\bf Z}$}}

\newcommand{\cp} {\mbox{${\bf CP^{1}}$}}

\newcommand{\Flm}{\mbox{${\cal F}_{\lambda\mu}$}}

\newcommand{\calp}{\mbox{${\cal P}$}}

\newcommand{\ca}{\mbox{${\cal A}$}}

\newtheorem{proposition}{Proposition}[section]

\newtheorem{theorem}{Theorem}[section]

\newtheorem{lemma}[theorem]{Lemma}
\title{{ \bf Fusion algebra at a rational level and cohomology
of nilpotent subalgebras of $\widehat{\gtsl_{2}}$   }}
\author{Boris Feigin\\
Landau Institute for Theoretical Physics
 \and
Feodor Malikov\\
 Department of Mathematics, Yale University }
\date{Received: }

\begin{document}

\maketitle

\begin{abstract}
We define and calculate the fusion algebra of WZW model at a rational
level by cohomological methods. As a byproduct we obtain a
cohomological
characterization of admissible representations of $\widehat{\gtsl}_{2}$.
\vspace{5 mm}

\end{abstract}

\section{{\bf Introduction}}
The purpose of this note is to define and calculate the fusion algebra
of WZW model at a rational level using cohomological methods. The link
between
cohomology theory of infinite dimensional Lie algebras
and conformal field theory was observed quite some time ago. It is
provided
by the construction which attaches algebras in question ( usually
Virasoro  or an affine Lie algebra) and modules over them to a point on
a curve.
In a sense, this makes a module depend on the additional parameter,
that is a
point on a curve.
 This construction gives satisfactory results in the Virasoro
algebra
case (~\cite{feig_fuchs}).
 It does also in the affine Lie algebra case ( ~\cite{ts_u_ya}) provided the
theory is related to an integral highest weight.
( An impressive application of this technique
is to be found in ~\cite{kazh_luszt} .) Several observations
indicate, however, that difficulties arise in the rational highest
weight
case. Among these we mention that the calculation
of the fusion algebra based on Verlinde
formula
gives negative structure constants
 ( ~\cite{mat_walt} ), the result lacking natural
interpretation.

A partial solution to the problem
 was given by Awata and Yamada  ~\cite{awata} , where vertex operators
( intertwiners between certain affine Lie algebra
modules, for the definition
see
the main body of paper) were classified. This allowed to envisage what
the structure constants of the fusion algebra are. What was still
lacking
was the rich structure of conformal field theory, which apart from
vertex
operators possesses operators of a more complicated nature and, quite
remarkably, assigns an element of a highest weight module to each
operator so that, for example, a vertex operator is associated with a
highest weight vector.

Here we propose an improvement of the construction attaching a module
to
a point. We suggest that one more parameter should be introduced and a
module
should be attached to a pair `` point on a curve, Borel subalgebra of the
underlying finite dimensional Lie algebra which the module is induced
from''.
After that the usual definition of the conformal block in spirit of
{}~\cite{feig_fuchs,ts_u_ya}
goes through nicely. In particular, in the case of
$\widehat{\gtsl}_{2}$ on $\cp$ one reconstructs the above-mentioned
structure
of conformal field theory and calculates the fusion algebra,
recovering
the result of Awata and Yamada.

Our calculation of the fusion algebra rests on the calculation of the
cohomology
of a certain nilpotent subalgebra $\gta\subset\widehat{\gtsl}_{2}$.
The latter might be of some interest in its own right. It is known
that the infinite dimensional Lie algebra analogues of finite
dimensional
irreducible modules over a simple Lie algebra are infinite
dimensional.
Such are minimal representations of Virasoro algebra and integrable
highest
weight modules, i.e. irreducibles with dominant integral highest
weight,
over an affine Lie algebra. It is shown in ~\cite{feig_fuchs} that,
nevertheless,
these
modules are singled out by a certain finiteness condition: they are
equivalently described as all modules $V$ such that
the 0th cohomology of any
finite codimension nilpotent subalgebra with coefficients in $V$ is
finite
dimensional. This completes the Virasoro algebra case. In the affine
case,
however, there is a broad class of modules containing integrable
modules as a subclass. Modules of this class were called
admissible by Kac and
Wakimoto
{}~\cite{kac_wak}). In many respects they are no worse than the integrable; in
particular their characters are modular invariant. Their consideration
is also necessitated by the quantum hamiltonian reduction
( ~\cite{f_g_p_p} and references therein ), i.e. by the
functor between the categories of Virasoro and $\widehat{\gtsl}_{2}$
modules ( ~\cite{feig_fr}).
This functor sends highest weight modules to highest weight modules
and
establishes 1-1 correspondence between minimal representations of
Virasoro
algebra and admissible modules over  $\widehat{\gtsl}_{2}$.

Our calculation shows that admissible modules are singled out
in the class of modules at the level $k> -2$ by a
similar
condition: they are exactly such modules $V$ that
$dim\, H^{0}(\gta,V)<\infty$.

Part of the above is easy to extend to the higher rank case, part is
not quite so. We intend to address this issue in the forthcoming paper.

{\bf Acknowledgments} Part of the work was done when F.M. was visiting
Department of Applied Mathematics and Theoretical Physics of
University
of Cambridge supported by SERC Fellowship. F.M. wishes to thank his
colleagues from DAMTP for hospitality and many interesting discussions.

\section{{\bf Notations and known results}}
\label{notat_known_res}

1.Set $\gtg = \gtsl_{2}$, $\hgtg=\widehat{\gtsl}_{2}=\gtsl_{2}
\otimes\nc [z,z^{-1}]\oplus\nc c$. Choose a basis $e,\,h,\,f$ of
$\gtg$
satisfying $[h,e]=2e,\;[h,f]=-2f,\;[e,f]=h$. $\gtb=\nc e\oplus \nc h$
and
$\hgtb =\gtg\otimes z\nc[z]\oplus \gtb\oplus\nc c$ are standard Borel
subalgebras of $\gtg$ and $\hgtg$ resp.

The Verma module $M_{\lambda,k}$ is a module induced from the
character
of $\gtg\otimes z\nc[z]\oplus \gtb\oplus\nc c$ annihilating
$\gtg\otimes z\nc[z]\oplus \nc e$ and sending $ h$ and $c$ to
$\lambda$ and $k$ resp. $k$ is often referred to as
a level. Generator of $M_{\lambda,k}$ is usually denoted by
$v_{\lambda,k}$.

The algebra $\hgtg$ is $\nz^{2}_{+}-$graded by assigning $f\otimes
z^{n}\mapsto
(1,-n),\;e\otimes z^{n}\mapsto (-1,-n)$ and so is
$M_{\lambda,k}=\oplus
_{i,j}M_{\lambda,k}^{i,j}$.

A morphism of Verma modules $M_{\lambda,k}\rightarrow M_{\mu,k}$ is
determined by the image of $v_{\lambda,k}$. It can be written as
$Sv_{\mu,k}$ for a uniquely determined element $S$ of the universal
enveloping algebra of  $\gtg\otimes z^{-1}\nc[z^{-1}]\oplus\nc f$.
Vector $Sv_{\mu,k}$ or even $S$ for this matter is called singular.

The  structure of Verma modules over $\hgtg$ is known in full detail
(\cite{mal_2}).
Outside the critical level ($k=-2$) and apart from more trivial possibilities,
$M_{\lambda,k}$ contains 2 independent singular vectors and these
generate
the maximal proper submodule. In the region $k>-2$ a Verma module
contains
infinitely many singular vectors and is embedded in finitely many
other
Verma modules ( the situation in the region $k<-2$ is dual to this).
Although formally all such Verma modules look alike a special role is
played
by those which can only embed (non-trivially) in themselves. Highest
weights of such modules are
called admissible (~\cite{kac_wak}) and are described as follows:

Let $k+2=p/q$, where $p,q$ are relatively prime positive integers.
The set of admissible highest weights at the level $k=p/q-2$ is given
by
\[\Lambda_{k}=\{\lambda(m,n)=m\frac{p}{q}-n-1\, :\;0<m\leq q,\,
0\leq n\leq p\}.\]

What is said above about the structure of Verma modules implies that
any
Verma module appears in the exact sequence of the form
\begin{equation}
\label{bgg_res}
0\leftarrow L_{\lambda_{0},k}\leftarrow
M_{\lambda_{0},k}\stackrel{d_{0}}{\leftarrow} M_{\lambda_{1},k}\oplus
M_{\mu_{1},k}
\stackrel{d_{1}}{\leftarrow} M_{\lambda_{2},k}\oplus
M_{\mu_{2},k}\stackrel{d_{2}}{\leftarrow}\cdots,
\end{equation}
where $\lambda_{0}$ is an admissible weight at the level $k$ and
$L_{\lambda_{0},k}$ is
the corresponding irreducible module.
$L_{\lambda_{0},k}$ is also called admissible.
 The exact sequence
(~\ref{bgg_res}) is called Bernstein - Gel`fand - Gel'fand ( BGG ) resolution.

2. {\em Sigular vector formula.}
 It follows from Kac-Kazhdan
determinant formula that a singular vector generically appears in the
homogeneous
components
of degree either $N(-1,l),\;l> 0,\,N>0$ or $N(1,l),\;l\geq 0,l\geq 0$.
Denote
the corresponding singular vectors by $S_{N,l}^{1}$ and
$S_{N,l}^{2}$ resp.

 Singular vectors $S_{Nl}^{i}$ were found in ~\cite{malff}
in an unconventional form containing  non-integral powers of
elements
of $\hgtg$ ( see also ~\cite{ba_soch} for another approach):
\begin{eqnarray}
S_{Nl}^{1}=(e\otimes z^{-1})^{N+lt}f^{N+(l-1)t}
(e\otimes z^{-1})^{N+(l-2)t}\cdots (e\otimes z^{-1})^{N-lt}
\label{s_v_1}\\
S_{Nl}^{2}=f^{N+lt}(e\otimes z^{-1})^{N+(l-1)t}
f^{N+(l-2)t}\cdots f^{N-lt}
\label{s_v_2},
\end{eqnarray}
where $t=k+2$.

This form is not always convenient to calculate a singular vector. It is,
however,
 a useful tool to derive properties of a singular vector. For example,
denoting
by
$\pi:\hgtg\rightarrow\gtg,\;g\otimes z^{n}\mapsto g$ the evaluation
map,
one uses (~\ref{s_v_1},~\ref{s_v_2}) to derive that (see ~\cite{fuchs}, also
{}~\cite{mal} for
the proof in a more general quantum case):

\begin{eqnarray}
\pi S_{Nl}^{1}=(\prod_{i=1}^{l}\prod_{j=1}^{N} P(-it-j))e^{N}
\label{p_s_v_1}\\
\pi S_{Nl}^{2}=(\prod_{i=1}^{l}\prod_{j=0}^{N-1} P(it+j))f^{N}
\label{p_s_v_2},
\end{eqnarray}
where $P(t)=ef-(t+1)h-t(t+1)$.

 3. We will also be using $\hgtg-$modules different from Verma modules
or corresponding irreducible ones:

Denote by $M_{\lambda,k}^{opp}$ a module induced from the Borel
subalgebra
opposite to $\hgtb$; obviously
 $M_{\lambda,k}^{opp}$ is a lowest weight module. The canonical
automorphism
of $\hgtg$ determined by: $e\otimes z^{n}\mapsto f\otimes
z^{-n},\;f\otimes
z^{n}\mapsto
e\otimes z^{-n}$ shows that $M_{\lambda,k}$ and $M_{\lambda,k}^{opp}$
have
formally identical properties. For example,  both are
$\nz_{+}\times\nz_{+}$-graded, singular vectors appear in the same
homogeneous
components and are given by

\begin{eqnarray}
T_{Nl}^{1}=(f\otimes z)^{N+lt}e^{N+(l-1)t}
(f\otimes z)^{N+(l-2)t}\cdots (f\otimes z)^{N-lt}
\label{s_v_1_opp}\\
T_{Nl}^{2}=e^{N+lt}(f\otimes z)^{N+(l-1)t}
e^{N+(l-2)t}\cdots e^{N-lt}
\label{s_v_2_opp}.
\end{eqnarray}

Observe that
\begin{equation}
\label{pr_s_v_opp}
\pi T_{Nl}^{1}=\pi S_{Nl}^{2};\pi T_{Nl}^{2}=\pi S_{Nl}^{1}
{}.
\end{equation}

 Set
$M_{\lambda,k}^{c}=
 (M_{\lambda,k}^{opp})^{\ast}$;  $M_{\lambda,k}^{c}$ is called
contragredient
Verma module.

Denote by $\Flm $ a $\gtg-$ module with the basis $F_{i},\;i\in\nz$
and the action given by
\[eF_{i}=-(\mu+i)F_{i-1},\;hF_{i}=(-2\mu-2i+\lambda)F_{i},\;
fF_{i}=(\mu+i-\lambda)F_{i+1}.\]
The space $\Flm [z,z^{-1}]=\Flm\otimes\nc [z,z^{-1}]$ is endowed with
the natural $\hgtg-$module structure. The elements $F_{ij}=
F_{i}\otimes z^{-j},\;i,j\in\nz$
serve as a natural basis in it.

4. Some notations from the commutative algebra are as follows:

$\nc [t]$ is a polynomial ring, $\nc [[t]]$ is its completion by
positive powers of $t$; $\nc[t,t^{-1}]$ is a ring of Laurent
polynomials and $\nc ((t))$ is its completion by positive powers of $t$.

\section{{\bf Fusion algebra of WZW model at a rational level}}
1.  From now on $\gtg$ stands for $\gtsl_{2}$. For any $w\in\cp$ define
$\gtg_{w}$
to be the algebra of functions defined over the
``punctured formal neighborhood of
$w$''
with values in $\gtg$. By this we mean that for any local coordinate $z$ in
 a neighborhood
of $w$ one has $\gtg_{w}\approx \gtg\otimes\nc((z-w))$.
$\gtg_{w}$ is equipped with a natural structure of a Lie algebra. It
possesses
the unique ( up to proportionality)
1-dimensional central extension determined by
the
cocycle $<f(z),g(z)>_{w}=Res_{z=w}Tr\,f(z)'g(z)\,dz$. Denote
this central extension by $\hgtg_{w}$.
The choice
of a cocycle fixes a central element $c$ so that $\hgtg_{w}=\gtg_{w}\oplus\nc
c$.
Obviously  $\hgtg = \hgtg_{0}$.

Further, for a finite $E\subset\cp$ set $\gtg_{E}=\oplus_{w\in
E}\gtg_{w}$ and define $\hgtg_{E}$ the 1-dimensional central extension
of $\gtg_{E}$ determined by he cocycle $<.,.>_{E}=\sum_{w\in E}<.,.>_{w}$.

For any finite  $E\subset\cp$ denote by $\gtg (E)$ the Lie
algebra
of meromorphic functions on $\cp$ and holomorphic outside $E$ with
values in $\gtg$. The Taylor expansions provide a homomorphism
\begin{equation}
\gtg(E)\rightarrow \gtg_{E}.
\end{equation}
It is less obvious but follows from the residue theorem that the above
homomorphism lifts to the homomorphism
\begin{equation}
\label{imb_in_dir_sum}
\gtg(E)\rightarrow \hgtg_{E}.
\end{equation}

2. It is natural to refer to $\gtg_{w}$ ($\hgtg_{w}$ resp.) as a current
algebra
( an affine Lie algebra resp. ) attached to the point $w\in \cp$.
In a similar manner one attaches a highest weight $\hgtg-$module to a point
$w\in\cp$.  Fix a pair of complex numbers $\lambda, k$. Observe that
$\cp$ is a flag manifold for $\gtg$. This means that there is a
canonical Borel subalgebra $\gtb_{w}\subset\gtg$ associated to any
$w\in\cp$. Fix Cartan generators of $h_{w},e_{w}\in \gtb_{w}$
satisfying
$e_{w}\in [\gtb_{w},\gtb_{w}],\;[h_{w},e_{w}]=2e_{w}$.
This determines a character $\phi_{\lambda}:\;\gtb_{w}\rightarrow \nc$
satisfying $\phi_{\lambda}(e_{w})=0,\;\phi_{\lambda}(h_{w})=\lambda$.
Although the choice of the generators $h_{w},e_{w}$ is not quite
unique,
the character $\phi_{\lambda}$ only depends on $\lambda\in\nc$.

Set $\hat{\gtb}_{w}=\gtb_{w}\oplus(\gtg\otimes(z-w)\nc [[z-w]])\oplus \nc
c$. Making use of the other complex number $k$ one extends
$\phi_{\lambda}$ to the character $\phi_{\lambda,k}$  of $\hat{\gtb}_{w}$
by
setting\[\phi_{\lambda,k}|_{\gtg\otimes(z-w)\nc [[z-w]]}=0,\;
\phi_{\lambda,k}(c)=k.\]

{\em A Verma module $M_{\lambda,k}(w)$ attached to $w\in\cp$} is said
to be a $\hgtg_{w}$-module induced from the character
$\phi_{\lambda,k}$
of $\hat{\gtb}_{w}$. Obviously $M_{\lambda,k}(0)$ is the usual Verma
module
$M_{\lambda , k}$, while $M_{\lambda,k}(\infty)$ is the opposite Verma
module
$M_{\lambda,k}^{opp}$, see sect.~\ref{notat_known_res}.
Taking a  quotient of
$M_{\lambda,k}(w)$ one obtains the definition of an arbitrary highest
weight module
$V_{\lambda,k}(w)$ attached to $w$. Denote by $L_{\lambda,k}(w)$ the
corresponding
irreducible highest weight module.

3. By the definition, the tensor product
$V_{\lambda_{1},k}(w_{1})\otimes\cdots\otimes V_{\lambda_{m},k}(w_{m})$ is
a $\hgtg_{E}-$module, where $E=\{w_{1},\ldots ,w_{m}\}$.
 The pull-back with respect to the homomorphism
(\ref{imb_in_dir_sum}) makes it  a $\gtg (w_{1},\ldots
,w_{m})-$module. We will be interested in the space of coinvariants
\[(V_{\lambda_{1},k}(w_{1})\otimes\cdots\otimes V_{\lambda_{m},k})^
{\gtg(E)}=H^{0}(\gtg(E),\;
V_{\lambda_{1},k}(w_{1})\otimes\cdots\otimes V_{\lambda_{m},k})\]
for appropriate $V_{\lambda_{i},k},\;1\leq i\leq m$ and
and $m$.

\begin{lemma}
\label{cornerstone}
Let $w_{1},w_{2},w_{3}\in\cp$ be 3 distinct points. Then
\[(i)\;\;dim\, H^{0}(\gtg(w_{1},w_{2}),\;
L_{\lambda_{1},k}(w_{1})\otimes L_{\lambda_{2},k}(w_{2}))=
\left\{\begin{array}{ll}
         1 & \mbox{ if } \lambda_{1}=\lambda_{2}\\
         0 & \mbox{ otherwise. }
       \end{array}\right.\]
\[(ii)\;\;dim\, H^{0}(\gtg(w_{1},w_{2},w_{3}),\;
M_{\lambda_{1},k}(w_{1})\otimes M_{\lambda_{2},k}(w_{2})
\otimes M_{\lambda_{3},k}(w_{3}))=1\]
for any $\lambda_{1},\lambda_{2},\lambda_{3}$.
\end{lemma}

4. The relation of the above to physics is as follows. A conformal
field
theory is ( in particular ) a collection of irreducibles
$L_{\lambda,k}$ for
 a fixed $k$ and $\lambda$ running over an appropriate set $\Lambda$.
The space $H^{0}(\gtg(w_{1},\ldots , w_{m}),\;
L_{\lambda_{1},k}(w_{1})\otimes\cdots\otimes L_{\lambda_{m},k}(w_{m}))$ is
called
conformal block. The fusion algebra is also defined in terms of
coinvariants.

Let $\ca$ be a vector space with the basis
$\{l_{\lambda},\;\lambda\in\Lambda\}$. This space carries a family of
symmetric forms $\Phi_{m},\;m\geq 2$ defined by
\[\Phi_{m}(l_{\lambda_{1}},\ldots ,l_{ \lambda_{m}})=
dim\,H^{0}(\gtg(w_{1},\ldots , w_{m}),\;
L_{\lambda_{1},k}(w_{1})\otimes\cdots\otimes L_{\lambda_{m},k}(w_{m})).\]
Item (i) of Lemma ~\ref{cornerstone} implies that $\Phi_{2}$ provides
an isomorphism $\ca^{\ast}\approx\ca$ and the basis $\{l_{\lambda},\;\lambda
\in\Lambda\}$ is self-dual. Therefore $\Phi_{3}$ determines a
commutative
linear map $\ca\otimes\ca\rightarrow\ca$. $\ca$ with thus defined
commutative algebra structure is called fusion algebra. Here we
calculate
this algebra structure in the case when $\Lambda=\Lambda_{k}$ is the set
of all admissible highest weights at a rational level $k$.

5. {\bf Main result.}
Let $k+2=p/q$, where $p,q$ are relatively prime positive integers.
The set of admissible highest weights at the level $k=p/q-2$ is given
by
\[\Lambda_{k}=\{\lambda(m,n)=m\frac{p}{q}-n-1\, :\;0<m\leq q,\,
0\leq n\leq p-1\}.\]
We will call a triple of admissible weights
$\lambda_{i}=\lambda(m_{i},n_{i}),\; i=1,2,3$ {\em proper} if it
satisfies either
\[\left\{\begin{array}{lll}
 max\{ 2+m_{3}-m_{1},2+m_{1}-m_{3}\}\leq & m_{2} &\leq min
\{ 2q-m_{3}-m_{1},m_{3}+m_{1}-2\}\\
max\{ n_{3}-n_{1}+1,n_{1}-n_{3}+1\}\leq & n_{2} & \leq
min\{ n_{1}+n_{3}-1,-n_{1}-n_{3}+2p-1\}
       \end{array}\right.\]

or
\[\left\{\begin{array}{lll}
 max\{ m_{3}+m_{1}-q,-m_{3}-m_{1}+q+2\}\leq & m_{2} &\leq min
\{ m_{1}-m_{3}+q,m_{3}-m_{1}+q\}\\
max\{ n_{3}+n_{1}-p+1,-n_{1}-n_{3}+p+1\}\leq & n_{2} & \leq
min\{ n_{1}-n_{3}+p-1,n_{3}-n_{1}+p-1\}
       \end{array}\right. \]
where it is understood that in both cases $m_{2},n_{2}$ run between
the limits with the step equal to 2; note that in all the cases the
limits
are of the same parity.

Denote by $\calp _{k}$ the set of proper weights at the level $k$.

\begin{theorem} The algebra structure of $\ca$ at the level $k=p/q-2$
is
given by
\[l_{\lambda_{0}}\cdot l_{\lambda_{1}}=\sum_{\lambda:
(\lambda_{0},\lambda_{1},\lambda)\in \calp
_{k}}\lambda.\]
\label{main_theor}
\end{theorem}
Proof of this theorem is based on calculation of cohomology of a
certain
subalgebra of $\hgtg$ with coefficients in a highest weight module
(see sect.~\ref{coh_of_nilp_subalg}).

6. We finish this section by describing how one obtains vertex
operators, i.e. $\hgtg-$ morphisms
\begin{equation}
\label{def_vert_oper}
\Flm [z,z^{-1}]\otimes M_{\lambda_{1},k}\rightarrow M_{\lambda_{2},k}^{c}
\end{equation}
through coinvariants. It is best understood by generalizing
the above construction of $M_{\lambda , k}(z)$ to obtain
a module $M_{\lambda , k}(w_{1},w_{2})$ attached to the point
$(w_{1},w_{2})\in\cp\times\cp$. The last module is defined
to be induced from the character of the subalgebra
\[(\gtg\otimes (z-w_{1})\nc[[z-w_{1}]])\oplus\gtb_{w_{2}}\oplus
\nc c\]
vanishing on $(\gtg\otimes (z-w_{1})\nc[[z-w_{1}]])\oplus \nc
e_{w_{2}}$
and sending $h_{w_{2}},\;c$ to $\lambda,\;k$ resp. It is clear that
$M_{\lambda , k}(w,w)=M_{\lambda , k}(w)$.

The following analogue of Lemma~\ref{cornerstone} (ii) takes place:
\[(ii)\;\;dim\, H^{0}(\gtg(0,w_{1},\infty),\;
M_{\lambda_{1},k}(0)\otimes M_{\lambda_{2},k}(w_{1},w_{2})
\otimes M_{\lambda_{3},k}(\infty))=1\]
for any $\lambda_{1},\lambda_{2},\lambda_{3}$ ( we have fixed
some of the parameters ). Let $v_{0},v(w_{1},w_{2}),v_{\infty}$
be highest weight vectors of the modules
$M_{\lambda_{1},k}(0),\, M_{\lambda_{2},k}(w_{1},w_{2}),\,
 M_{\lambda_{3},k}(\infty)$ resp. Normalizing a cocycle generating
\[H^{0}(\gtg(0,w_{1},\infty),\;
M_{\lambda_{1},k}(0)\otimes M_{\lambda_{2},k}(w_{1},w_{2})
\otimes M_{\lambda_{3},k}(\infty))\] so that it is equal to 1 on
$v_{0}\otimes v(w_{1},w_{2})\otimes v_{\infty}$ one realizes
$v(w_{1},w_{2})$ as an element of the dual space
$(M_{\lambda_{1},k}(0)\otimes M_{\lambda_{3},k}(\infty)^{\ast}$. Observe that
the definitions imply the $\hgtg-$isomorphism:
$M_{\lambda_{3},k}(\infty)^{\ast}\approx M_{\lambda_{3},k}^{c}$, where
$M_{\lambda_{3},k}^{c}$ stands for the contragredient Verma module.
This means that the Fourier components of the operator function
$v(w_{1},w_{2})=\sum_{i,j}v^{-i,-j}w_{1}^{i}w_{2}^{j}$ are mappings
\[v^{ij}:\; M_{\lambda_{1},k}\rightarrow M_{\lambda_{3},k}^{c}\]
of degree $(-i,-j)$:
\[v^{ij}(M_{\lambda_{1},k}^{mn})\subset
(M_{\lambda_{3},k}^{c})^{m-i,m-j},\]
for all $m,n$. The algebra $\hgtg$ naturally acts on
$v^{ij},\;i,j\in\nz$
as elements of $Hom(M_{\lambda_{1},k},M_{\lambda_{3},k}^{c})$. The
following is proved by direct calculations using the definitions.

\begin{proposition}
\[\oplus_{i,j\in\nz}\nc v^{ij}\approx \Flm[z,z^{-1}],\]
where
$\lambda=-\lambda_{2}-2,\;\mu=(\lambda_{1}-\lambda_{2}-\lambda_{3}-2)/2.$
\end{proposition}
Therefore the highest weight vector $v(w_{1},w_{2})$ gives rise to a
vertex operator (see (~\ref{def_vert_oper}) ). Further, other elements
of $M_{\lambda_{2},k}(w_{1},w_{2})$ (``descendants of the vacuum'')
give rise
in a similar way to more complicated $\hgtg$-modules realized in
$Hom(M_{\lambda_{1},k},M_{\lambda_{3},k}^{c})$. This is one of the
main structures of WZW model.

\section{{\bf Cohomology of a nilpotent subalgebra of $\hgtg$.}}
\label{coh_of_nilp_subalg}
 1. Set $\gta = \gtg\otimes z(z-1)\nc [z]\oplus \nc e\otimes(z-1)
\oplus \nc (e-h-f)\otimes z$,
$\bar{\gta}=\gta\oplus \nc h\otimes (z-1)\oplus \nc (h+2f)\otimes z$.
Obviously, $\gta$ and $\bar{\gta}$ are commutative subalgebras of
$\hat{\gtb}$. Moreover, $\gta\subset\bar{\gta}$ is an ideal,
$\bar{\gta}/\gta$ is a 2-dimensional commutative algebra generated
by the classes of $h\otimes (z-1)$ and $(h+2f)\otimes z$. This means
that
for any $\bar{\gta}$-module $V$
the elements  $h\otimes (z-1)$ and $(h+2f)\otimes z$ act on
 the cohomology groups $H^{r}(\gta , V)$ commuting with each other.

\begin{theorem}
 \label{main_th_cohom}

(i) $dim\, H^{0}(\gta , L_{\lambda , k}(\infty))<\infty$ if and only if
$\lambda$ is admissible.

(ii) Set $\lambda = \lambda (m,n)$.  Operators
 $h\otimes (z-1)$, $(h+2f)\otimes z$  have simple spectra in
$H^{0}(\gta , L_{\lambda ,k}(\infty))$. If
$H^{0}(\gta , L_{\lambda ,k}(\infty))^{(\alpha , \beta)}$ is a simultaneous
eigenspace, then

\[dim\, H^{0}(\gta , L_{\lambda,k}(\infty))^{(\alpha,\beta)}=
\left\{\begin{array}{ll}
         1 & \mbox{ if } (\lambda,-\alpha,\beta) \mbox{ is proper }\\
         0 & \mbox{ otherwise. }
       \end{array}\right.\]
\end{theorem}
{\em Remark.} As explained above, attaching an irreducible highest
weight module
to $\infty$ means changing it to the irreducible lowest weight module.
We use such a module in the last theorem for purely technical reasons.
The
result is immediately interpreted in terms of highest weight modules.

2. Sketch of the  proof of  Theorem ~\ref{main_th_cohom}.
A Verma module attached to $\infty$ is $\gta-$free. Therefore
cohomology of $\gta$ can be calculated by means of BGG resolution
(\ref{bgg_res}) or, rather, its analogue for lowest weight modules. In
particular,
the 0th cohomology is equal to the kernel of $d_{0}^{\ast}$ in
$(M_{\lambda,k}(\infty)^{\ast})^{\gta}$. $M_{\lambda,k}(\infty)$ is
$\nz_{+}\times
\nz_{+}$-graded, therefore, with an element $F\in
M_{\lambda,k}(\infty)^{\ast}$ one can associate
fuctionals $F_{ij},\,i,j\geq 0$, where $F_{ij}$ is a restriction of
$F$ to the homogeneous subspace of degree $(i,j)$.
\begin{lemma}(\mbox{c.f. Proposition 3.1})
 If $h\otimes (z-1)F=\alpha F,\;(h+2f)\otimes z F=\beta
F$ then
with respect to the natural action
of $\gtg\otimes z\nc[z]\oplus \nc e$, the space
 $\oplus_{i,j\geq 0}\nc F_{ij}$ is isomorphic with
$\Flm$ for $\lambda=-\beta-2,\;\mu=-1/2(\lambda +\alpha + \beta)-1$.
\end{lemma}

$M_{\lambda,k}(\infty)$ contains 2 singular vectors, say $S_{1}$ and
$S_{2}$,
of degrees $(r,s)$ and $(r_{1},s_{1})$ resp. Therefore, $ker\,
d_{0}^{\ast}$
is given by the following system
\begin{equation}
\left\{\begin{array} {rr} S_{1}F_{rs}=0\\
S_{2}F_{r_{1}s_{1}}=0\end{array}
\right.\end{equation}

Action of $\hgtg$ on elements of $\Flm$ is given through the
evaluation map.
Formula (~\ref{pr_s_v_opp}) shows that each of the above equations
determines a collection of straight lines in $(\alpha,\beta)-$plane.
Calculations show that if $\lambda$ is not admissible then the 2
families have at least 1 common line, proving (i). If $\lambda$ is
admissible then the same calculations give (ii).

3. Here we will derive Theorem ~\ref{main_theor} from
 Theorem ~\ref{main_th_cohom}. Definitions imply that the claim of
Theorem ~\ref{main_theor} is equivalently reformulated as follows
\begin{equation}
\label{prelim_1}
H^{0}(\gtg(0,1,\infty),\;L_{\lambda_{1},k}(0)\otimes
L_{\lambda_{2},k}(1)
\otimes L_{\lambda_{3},k}(\infty))=
\left\{\begin{array}{ll}
         \nc & \mbox{ if } (\lambda_{1},\lambda_{2},\lambda_{3})
 \mbox{ is proper }\\
         0 & \mbox{ otherwise. }
       \end{array}\right.
\end{equation}

First, prove that the last formula holds for Verma modules, i.e.

\begin{equation}
\label{prelim_2}
H^{0}(\gtg(0,1,\infty),\;M_{\lambda_{1},k}(0)\otimes
M_{\lambda_{2},k}(1)
\otimes L_{\lambda_{3},k}(\infty))=
\left\{\begin{array}{ll}
         \nc & \mbox{ if } (\lambda_{1},\lambda_{2},\lambda_{3})
 \mbox{ is proper }\\
         0 & \mbox{ otherwise. }
       \end{array}\right.
\end{equation}
To do so observe that the module
$M_{\lambda_{1},k}(0)\otimes
M_{\lambda_{2},k}(1) $ is induced
from the ``2- character'' of $\bar{\gta}$ determined by the composition map
\[\bar{\gta}\rightarrow \bar{\gta}/\gta\rightarrow\nc^{2},\]
where $h\otimes (z-1)\mapsto -\lambda_{1},\; (h+2f)\otimes
z\mapsto\lambda_{2}.$
Frobenius duality gives that
\begin{eqnarray}
H^{0}(\gtg(0,1,\infty),\;M_{\lambda_{1},k}(0)\otimes
M_{\lambda_{2},k}(1)
\otimes L_{\lambda_{3},k}(\infty))=\nonumber\\
H^{0}(\gta,\, L_{\lambda_{3},k}(\infty))^{(\lambda_{1},\lambda_{2})}.\nonumber
\end{eqnarray}
Equality (\ref{prelim_2}) now follows from Theorem
{}~\ref{main_th_cohom}.

To complete the proof of Theorem ~\ref{main_theor} it remains to show that
\begin{eqnarray}
H^{0}(\gtg(0,1,\infty),\;L_{\lambda_{1},k}(0)\otimes
L_{\lambda_{2},k}(1)
\otimes L_{\lambda_{3},k}(\infty))=\nonumber\\
H^{0}(\gtg(0,1,\infty),\;M_{\lambda_{1},k}(0)\otimes
M_{\lambda_{2},k}(1)
\otimes L_{\lambda_{3},k}(\infty)).\nonumber
\end{eqnarray}
 The following part of the BGG
resolution
\[M_{1}\oplus M_{2}\rightarrow M_{\lambda_{1},k}\rightarrow
L_{\lambda_{1},k}
\rightarrow 0\]
gives rise to the following exact sequence of the cohomology groups
\begin{eqnarray}
H^{0}(\gtg(0,1,\infty),\,(M_{1}\oplus M_{2})(0)\otimes
M_{\lambda_{2},k}(1) \otimes L_{\lambda_{3},k}(\infty)) \rightarrow
\nonumber\\
H^{0}(\gtg(0,1,\infty),\,(M_{\lambda_{1},k})(0)\otimes
M_{\lambda_{2},k}(1) \otimes L_{\lambda_{3},k}(\infty)) \rightarrow
\nonumber\\
H^{0}(\gtg(0,1,\infty),\,(L_{\lambda_{1},k})(0)\otimes
M_{\lambda_{2},k}(1) \otimes L_{\lambda_{3},k}(\infty))\rightarrow 0.
\nonumber\end{eqnarray}
Since the highest weights of $M_{1},\,M_{2}$ are not admissible
 see sect.~\ref{notat_known_res}, Theorem
{}~\ref{main_th_cohom} implies that
\[H^{0}(\gtg(0,1,\infty),\,(M_{1}\oplus M_{2})(0)\otimes
M_{\lambda_{2},k}(1) \otimes L_{\lambda_{3},k}(\infty))=0.\]
Therefore,
\begin{eqnarray}
H^{0}(\gtg(0,1,\infty),\,(M_{\lambda_{1},k})(0)\otimes
M_{\lambda_{2},k}(1) \otimes L_{\lambda_{3},k}(\infty)) =\nonumber\\
H^{0}(\gtg(0,1,\infty),\,(L_{\lambda_{1},k})(0)\otimes
M_{\lambda_{2},k}(1) \otimes L_{\lambda_{3},k}(\infty)).\nonumber
\end{eqnarray}
Similar arguments go through for the factor $M_{\lambda_{2},k}$.
Theorem
{}~\ref{main_theor} has been proved. $\Box$

\end{document}